\documentclass[	
	paper = a4,
	fontsize = 11pt,
	DIV = 12,
	abstract = true
]{scrartcl}
\addtokomafont{disposition}{\rmfamily}
\usepackage{authblk}
\usepackage[hidelinks]{hyperref}
\usepackage[utf8]{inputenc}

\usepackage{amsmath}
\usepackage{amssymb}

\usepackage{enumerate}
\usepackage{paralist}
\usepackage{tikz}
\usepackage{subfigure}

\begin{document}

\title{A note on block-and-bridge preserving maximum common subgraph algorithms
for outerplanar graphs}

\author{Nils M.\@ Kriege}
\author{Andre Droschinsky}
\author{Petra Mutzel}

\affil{\small Department of Computer Science\\
TU Dortmund University, Germany \\
\texttt{\{nils.kriege,andre.droschinsky,petra.mutzel\}@tu-dortmund.de}
}

\date{}

\maketitle


\begin{abstract}
Schietgat, Ramon and Bruynooghe~\cite{Schietgat2013} proposed a polynomial-time 
algorithm for computing a maximum common subgraph under the block-and-bridge 
preserving subgraph isomorphism (BBP-MCS) for outerplanar graphs.
We show that the article contains the following errors:
\begin{enumerate}[(i)]
 \item The running time of the presented approach is claimed to be 
  $\mathcal{O}(n^{2.5})$ for two graphs of order $n$. We show
  that the algorithm of the authors allows no better bound than $\mathcal{O}(n^4)$
  when using state-of-the-art general purpose methods to solve the matching 
  instances arising as subproblems.
  This is even true for the special case, where both input graphs are trees.
 \item The article suggests that the dissimilarity measure derived from BBP-MCS 
  is a metric. We show that the triangle inequality is not always satisfied and, 
  hence, it is not a metric.
  Therefore, the dissimilarity measure should not be used in combination with 
  techniques that rely on or exploit the triangle inequality in any way.
\end{enumerate}
Where possible, we give hints on techniques that are suitable to improve the 
algorithm.
\end{abstract}

\section{Introduction}
Graph comparison is getting increasingly important with the growth of data analysis tasks
on graphs and networks.
An important application occurs in molecular chemistry for the tasks of virtual screening of molecular data bases, substructure search of molecules, and the discovery of  structure-activity relationships within rational drug design.
Thereby, finding the largest substructure that two molecules have in common is a fundamental 
task~\cite{Kriege2019}. Since molecules can naturally be 
represented by graphs, the problem is phrased as maximum common subgraph problem, 
which is as follows.
Given two graphs, find a graph with a largest possible number of edges that is 
isomorphic to subgraphs of both input graphs.
This classical graph theoretical problem generalizes the subgraph isomorphism 
problem and is well-known to be $\mathsf{NP}$-hard in general graphs~\cite{Garey1979}.
Even deciding whether a forest~$G$ is isomorphic to a subgraph of a tree is an 
$\mathsf{NP}$-complete problem~\cite{Garey1979}. However,  if $G$ is a tree the
subgraph isomorphism problem can be solved in polynomial 
time~\cite{Matula1978,Reyner1977,Chung1987,Verma1989,Shamir1999}.
The generalisation of this approach to the maximum common subgraph problem is
attributed to J.~Edmonds~\cite{Matula1978}.
However, the vast amount of molecular graphs of interest are not trees, but 
outerplanar graphs, i.e., they admit a drawing on the plane without edge 
crossings such that all vertices are incident to the outer face of the drawing.
Even deciding whether a tree is isomorphic to a subgraph of an outerplanar graph
is $\mathsf{NP}$-complete~\cite{Syslo1982}. On the other hand, subgraph 
isomorphism can be solved in polynomial time when both graphs are biconnected and 
outerplanar~\cite{Lingas1989}.
More general, subgraph isomorphism can be solved in polynomial time in 
$k$-connected partial $k$-tree~\cite{Matousek1992,Gupta1994}.

Based on these theoretical findings, Horv\'{a}th, Ramon and Wrobel~\cite{Horvath2006}
proposed to consider so-called \emph{block-and-bridge-preserving} (BBP) subgraph 
isomorphism for mining frequent subgraphs in databases of outerplanar molecular 
graphs. The BBP subgraph isomorphism allows to consider blocks, i.e., the biconnected 
components, and the trees formed by the bridges separately and thereby can be 
solved in polynomial-time. Moreover, the approach yields chemical meaningful 
results, since it requires that the ring systems of molecules are preserved.

The maximum common subgraph problem in outerplanar graphs of bounded degree can
be solved in polynomial time~\cite{Akutsu2013}. Although molecular graph have 
bounded degree and are often outerplanar, the algorithm has a high running time 
and is probably not suitable for practical use.
Schietgat, Ramon and Bruynooghe~\cite{Schietgat2013} proposed to determine a 
maximum common subgraph under the BBP subgraph isomorphism and developed an algorithm 
with a claimed running time of $\mathcal{O}(n^{2.5})$ for two outerplanar graphs 
of order $n$. While the authors presented promising experimental results on graphs representing molecules, we show
that their theoretical analysis of their approach is flawed.
Moreover, we show that the proposed approach to derive a distance from the 
size (or weight) of the maximum common subgraph does not yield a proper metric.

\section{Preliminaries}

We briefly summarize the necessary terminology and notation.
A \emph{graph} $G=(V,E)$ consists of a finite set $V(G) = V$ of \emph{vertices}
and a finite set $E(G) = E$ of \emph{edges}, where each edge connects two distinct 
vertices. 
A \emph{path} of length $n$ is a sequence of vertices $(v_0, \dots, v_n)$ such 
that $\{v_{i}, v_{i+1}\} \in E$ for $0 \leq i < n$. 
A \emph{cycle} is a path of length at least $3$ with no repeated vertices except $v_0=v_n$.
A graph is \emph{connected} if there is a path between any two vertices.
A graph is \emph{biconnected} if for any two vertices there is a cycle containing them.
A \emph{tree} is a connected graph containing no cycles.
A graph $G$ with an explicit root vertex $r\in V(G)$ is called \emph{rooted} graph, denoted by $G^r$.
A graph $G'=(V',E')$ is a \emph{subgraph} of a graph $G=(V,E)$, written 
$G' \subseteq G$, if $V' \subseteq V$ and $E' \subseteq E$.
A \emph{block} is a maximal subgraph that is biconnected.
An edge is a \emph{bridge} if it is not contained in any block. 
A \emph{matching} in a graph $G$ is a subset of edges $M \subseteq E(G)$ such 
that no two edges in $M$ share a common vertex, i.e., 
$e \cap e' = \emptyset$ for all distinct edges $e, e' \in M$.
Given a bipartite graph $G$ with edge weights $w : E(G) \to \mathbb{R}$,
the \emph{weighted maximal matching problem} asks for a  matching 
$M \subseteq E$ in $G$ such that the weight $w(M) = \sum_{e \in M} w(e)$ is
maximal.\footnote{Note that in~\cite{Schietgat2013} matchings are defined as 
specific relations between sets, cf.~Definiton~15. The running time to compute a
matching then depends on the number of pairs with strictly positive weight. 
This can be expressed in a natural way by the number of edges in bipartite graphs.}

An \emph{isomorphism} between two graphs $G$ and $H$ is a bijection 
$\varphi : V(G) \to V(H)$ such that
$\{u,v\} \in E(G) \Leftrightarrow \{\varphi(u),\varphi(v)\} \in E(H)$ for all $u,v \in V(G)$.
We say that the edge $\{u,v\}$ is mapped to the edge $\{\varphi(u),\varphi(v)\}$ by $\varphi$.
A \emph{subgraph isomorphism} from a graph $G$ to a graph~$H$ is an isomorphism
between $G$ and a subgraph $H' \subseteq H$.
A graph $G$ is said to be \emph{subgraph isomorphic} to a graph~$H$, written 
$G \preceq H$, if a subgraph isomorphism from $G$ to $H$ exists.
A subgraph isomorphism from $G$ to $H$ is \emph{block and bridge preserving}
(BBP) if
\begin{inparaenum}[(i)]
 \item each bridge in $G$ is mapped to a bridge in $H$, and
 \item any two edges in different blocks in $G$ are mapped to different blocks in $H$.
\end{inparaenum}
We write $G \sqsubseteq H$ if a BBP subgraph isomorphism from $G$ to $H$ exists.
A (BBP) \emph{common subgraph} of two graphs $G$ and $H$ is a connected graph $I$ such that
$I \preceq G$ and $I \preceq H$ ($I \sqsubseteq G$ and $I \sqsubseteq H$). A (BBP)
common subgraph $I$ is \emph{maximum} w.r.t.\@ a weight function $w$ if 
there is no (BBP) common subgraph $I'$ with $w(I') > w(I)$. The two different
concepts, maximum common subgraph (MCS) and BBP-MCS, are illustrated in 
Figure~\ref{fig:example}.
The above definitions can be naturally extended to graphs with vertex and edge labels,
where an isomorphism must preserve labels and the weight function may depend on the labels.

\tikzstyle{vertex}=[circle, fill=black, scale=.6]
\tikzstyle{vertex2}=[circle, draw=black, thick, fill=white, scale=.6]
\tikzstyle{edge} = [draw,thick,-]
\begin{figure}
  \centering
  \subfigure[MCS $I$ with $w(I)=17$]{\label{fig:example:mcs}
\begin{tikzpicture}[auto,scale=.65]

	\foreach \name /\x / \y in {u1/-1.5-3.5/1.5, u2/0-3.5/1.5, u4/-1.5-3.5/0, u5/0-3.5/0}
        \node[vertex](\name) at (\x,\y) {};

	\phantom{\node[vertex2,label=right:{$v$}](u5) at (-3.5,0) {};}

	\draw[edge] (u1) -- (u2);
	\draw[edge] (u1) -- (u4);
	\draw[edge] (u2) -- (u5);
	\draw[edge,dashed] (u4) -- (u5);
	\draw[edge] (u1) -- (u5);

	\node[vertex](d1) at (-6.5,1.5) {};
	\node[vertex](d2) at (-8,0) {};
	\node[vertex](d3) at (-6.5,0) {};
	\node[vertex](v7) at (-8,1.5) {};
	\draw[edge,dashed] (d1) -- (d2);
	\draw[edge] (d1) -- (d3);
	\draw[edge] (d2) -- (d3);
	\draw[edge] (d1) -- (u1);
	\draw[edge] (d3) -- (u4);
	\draw[edge] (d1) -- (v7);

	\node (label) at (-5.7,-.6) {$G$};
\end{tikzpicture}\hspace{1cm}
\begin{tikzpicture}[auto,scale=.65]

	\foreach \name /\x / \y in {u1/-1.5-3.5/1.5, u2/0-3.5/1.5, u4/-1.5-3.5/0, u5/0-3.5/0}
        \node[vertex](\name) at (\x,\y) {};

	\phantom{\node[vertex2,label=right:{$v$}](u5) at (-3.5,0) {};}

	\draw[edge] (u1) -- (u2);
	\draw[edge] (u1) -- (u4);
	\draw[edge] (u2) -- (u5);
	\draw[edge] (u1) -- (u5);

	\node[vertex](d1) at (-6.5,1.5) {};
	\node[vertex](d2) at (-8,0) {};
	\node[vertex](d3) at (-6.5,0) {};
	\node[vertex](v7) at (-8,1.5) {};
	\draw[edge] (d1) -- (d3);
	\draw[edge] (d2) -- (d3);
	\draw[edge] (d1) -- (u1);
	\draw[edge] (d3) -- (u4);
	\draw[edge] (d1) -- (v7);

	\node (label) at (-5.7,-.6) {$H$};
\end{tikzpicture}
}
\subfigure[BBP-MCS $I$ with $w(I)=10$]{\label{fig:example:bbp_mcs}
\begin{tikzpicture}[auto,scale=.65]

	\node[vertex](u1) at (-1.5-3.5,1.5) {};
	\node[vertex2,label=right:{$u$}](u2) at (0-3.5,1.5) {};
	\node[vertex](u4) at (-1.5-3.5,0) {};
	\node[vertex2,label=right:{$v$}](u5) at (-3.5,0) {};
	\draw[edge,dashed] (u1) -- (u2);
	\draw[edge] (u1) -- (u4);
	\draw[edge,dashed] (u2) -- (u5);
	\draw[edge,dashed] (u4) -- (u5);
	\draw[edge,dashed] (u1) -- (u5);

	\node[vertex](d1) at (-6.5,1.5) {};
	\node[vertex2](d2) at (-8,0) {};
	\node[vertex](d3) at (-6.5,0) {};
	\node[vertex](v7) at (-8,1.5) {};
	\draw[edge,dashed] (d1) -- (d2);
	\draw[edge] (d1) -- (d3);
	\draw[edge,dashed] (d2) -- (d3);
	\draw[edge] (d1) -- (u1);
	\draw[edge] (d3) node[above, xshift=-.4cm, yshift=-.1cm] {$e$} -- (u4);
	\draw[edge] (d1) -- (v7);

	\node (label) at (-5.7,-.6) {$G$};
\end{tikzpicture}\hspace{1cm}
\begin{tikzpicture}[auto,scale=.65]

	\node[vertex](u1) at (-1.5-3.5,1.5) {};
	\node[vertex2,label=right:{$u'$}](u2) at (0-3.5,1.5) {};
	\node[vertex](u4) at (-1.5-3.5,0) {};
	\node[vertex2,,label=right:{$v'$}](u5) at (-3.5,0) {};

	\draw[edge,dashed] (u1) -- (u2);
	\draw[edge] (u1) -- (u4);
	\draw[edge,dashed] (u2) -- (u5);
	\draw[edge,dashed] (u1) -- (u5);

	\node[vertex](d1) at (-6.5,1.5) {};
	\node[vertex2](d2) at (-8,0) {};
	\node[vertex](d3) at (-6.5,0) {};
	\node[vertex](v7) at (-8,1.5) {};
	\draw[edge] (d1) -- (d3);
	\draw[edge,dashed] (d2) --  (d3);
	\draw[edge] (d1) -- (u1);
	\draw[edge] (d3) node[above, xshift=-.4cm, yshift=-.1cm] {$e'$} -- (u4);
	\draw[edge] (d1) -- (v7);

	\node (label) at (-5.7,-.6) {$H$};
\end{tikzpicture}
}
\caption{Two graphs $G$, $H$ and their MCS~\subref{fig:example:mcs} and 
BBP-MCS~\subref{fig:example:bbp_mcs}, where $w(I)=|V(I)|+|E(I)|$. 
Dashed edges and blank vertices are not part of the common subgraph. Note that 
in Figure~\subref{fig:example:bbp_mcs} the vertex at the bottom left cannot be 
included since $e$ is part of a block in $G$ and $e'$ is a bridge in $H$. 
The two vertices $u$ and $v$ of $G$ cannot be added, since the triangle 
containing $u'$ and $v'$ constitutes a distinct block of $H$.}
  \label{fig:example}
\end{figure}
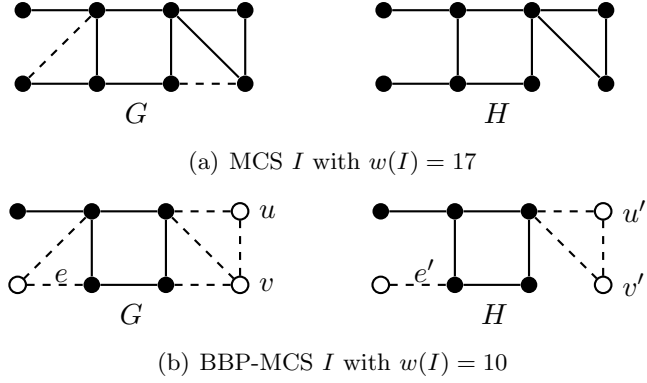

\section{Complexity Analysis}
The BBP-MCS algorithm for outerplanar graphs proposed in~\cite{Schietgat2013}
decomposes the two input graphs into subgraphs with distinct root vertices
referred to as \emph{parts} (see Section~\ref{sec:counterexample} for a formal 
definition).
An MCS problem for all compatible pairs of parts is then solved using a dynamic 
programming strategy.
Here, a series of weighted maximal matching instances arises as subproblems.
It has been claimed~\cite[Theorem~2]{Schietgat2013} that for two outerplanar 
graphs $G$ and $H$ the proposed BBP-MCS algorithm runs in time
\begin{equation*}\label{theorem2}
 \mathcal{O}\left(|V(G)| \cdot |V(H)| \cdot (|V(G)| + |V(H)|)^{\frac{1}{2}}\right),
\end{equation*}
which is $\mathcal{O}(n^{2.5})$ for $|V(G)| = |V(H)| = n$.
We show that this bound cannot be obtained by the presented techniques.

\subsection{Solving Weighted Maximal Matching Problems}\label{sec:mwbm}
The algorithm makes use of a subroutine for solving the weighted maximal matching
problem in bipartite graphs, where weights are real values.
The matching instances arising in the course of the algorithm may be complete 
bipartite graphs with a quadratic number of edges, see the counterexample 
discussed in Section~\ref{sec:counterexample}.
Hence, the running times given in the following refer to bipartite graphs with 
$n$ vertices and $\Theta(n^2)$ edges in order to improve readability.
The authors propose to use the algorithm by Hopcroft and Karp~\cite{Hopcroft1973} 
to solve an instance of the problem in time $\mathcal{O}(n^{2.5})$. Since this 
algorithm computes a matching of maximal cardinality, but is not designed to take 
weights into account, it cannot be applied to the instances that occur. 

The best known approaches for the weighted problem allow to solve instances with 
$n$ vertices and $\Theta(n^2)$ edges in time $\mathcal{O}(n^3)$, e.g., the 
established Hungarian method~\cite{Burkard2012}.
When we assume weights to be integers within the range of $[0..N]$,
scaling algorithms would become applicable such as~\cite{Duan2012}, which solves the
problem in time $\mathcal{O}(n^{2.5} \log N)$.
This running time is still worse than the time bound for the algorithm by Hopcroft 
and Karp by a factor depending logarithmically on $N$.
Moreover, it is desirable to allow that the weight of a common subgraph graph is 
measured by a real number depending on the labels of the vertices and edges it contains, 
cf.~\cite[Definition~2]{Schietgat2013}.
This leads to real edge weights in the matching instances.

In summary, no better bound than $\mathcal{O}(n^3)$ on the worst-case running time 
can be assumed for the subproblem of solving weighted maximal matching instances
with $n$ vertices.

\subsection{The Number of Matching Instances}\label{sec:counterexample}
We consider a particularly simple counterexample to illustrate that the running 
time required to solve the matching problems cannot be bounded by 
$\mathcal{O}(n^{2.5})$.
We identify the flaw regarding the analysis which led to this incorrect
result~\cite[Proof of Theorem~2]{Schietgat2013}.
More precisely, we show that for two graphs $G$ and $H$ of order $n$ the BBP-MCS 
algorithm performs $\Theta(n)$ calls to the subroutine for weighted maximal 
matching \cite[Algorithm 2, \textsc{MaxMatch}]{Schietgat2013} with instances of 
size $\Theta(n)$.
Since the relationship between the matching instances is not considered 
in~\cite{Schietgat2013}, we assume that each instance is solved separately in 
cubic time, cf. Section~\ref{sec:mwbm}. 
Therefore, no better bound than $\mathcal{O}(n^4)$ can be given on the total 
running time.

\begin{figure}
  \centering
  \null\hfill
  \subfigure[A star graph $H$]{\label{fig:star}
\begin{tikzpicture}[auto]
	\tikzstyle{vertex}=[circle, fill=black, scale=.6]
	\tikzstyle{edge} = [draw,thick,-]
	\tikzstyle{dots} = [line width=1.2pt, line cap=round, dash pattern=on 0pt off 3\pgflinewidth]

\def \n {5}
\def \radius {.8cm}
\def \margin {.8}

\node[vertex,label={right:$c$}] (r) at (0,0) {};
\node[vertex,label={left:$v_1$}] (v1) at ({115}:\radius) {};
\node[vertex,label={left:$v_2$}] (v2) at ({180}:\radius) {};
\node[vertex,label={left:$v_3$}] (v3) at ({245}:\radius) {};
\node[vertex,label={right:$v_4$}] (v4) at ({310}:\radius) {};
\node[vertex,label={right:$v_n$}] (vn) at ({50}:\radius) {};

\draw[edge] (r) -- (v1);
\draw[edge] (r) -- (v2);
\draw[edge] (r) -- (v3);
\draw[edge] (r) -- (v4);
\draw[edge] (r) -- (vn);

\draw[dots] ({330+\margin}:\radius) 
    arc ({330+\margin}:{390-\margin}:\radius);
\end{tikzpicture}
  }\hfill
  \subfigure[Rooted graph $H^c$]{\label{fig:star_root}
\begin{tikzpicture}[auto]
	\tikzstyle{vertex}=[circle, fill=black, scale=.6]
	\tikzstyle{edge} = [draw,thick,-]

\node[vertex,label ={above:$c$}]   (r) at (0,0) {};
\node[vertex,label ={below:$v_1$}] (v1) at (-1.02,-1) {};
\node[vertex,label ={below:$v_2$}] (v2) at (-.501,-1) {};
\node[vertex,label ={below:$v_3$}] (v3) at (-.0,-1) {};
\node[] (vdots) at (.5,-1) {$\cdots$};

\node[vertex,label ={below:$v_n$}] (vn) at (1.02,-1) {};

\draw[edge] (r) -- (v1);
\draw[edge] (r) -- (v2);
\draw[edge] (r) -- (v3);
\draw[edge] (r) -- (vn);

\phantom{\node[] () at (1.3,-1) {};
         \node[] () at (-1.3,-1) {};} 
\end{tikzpicture}
  }\hfill
  \subfigure[Part in $\mathcal{B}_H$]{\label{fig:elem}
\begin{tikzpicture}[auto]
	\tikzstyle{vertex}=[circle, fill=black, scale=.6]
	\tikzstyle{edge} = [draw,thick,-]

\node[vertex,label ={above:$c$}]   (r) at (0,0) {};
\node[vertex,label ={below:$v_1$}] (v1) at (-1.02,-1) {};
\node[vertex,label ={below:$v_2$},lightgray] (v2) at (-.501,-1) {};
\node[vertex,label ={below:$v_3$}] (v3) at (-.0,-1) {};
\node[] (vdots) at (.5,-1) {$\cdots$};

\node[vertex,label ={below:$v_n$}] (vn) at (1.02,-1) {};

\draw[edge] (r) -- (v1);
\draw[edge,lightgray] (r) -- (v2);
\draw[edge] (r) -- (v3);
\draw[edge] (r) -- (vn);

\end{tikzpicture}
  }\hfill\null
  \caption{\subref{fig:star} A star graph of order $n+1$, \subref{fig:star_root} the star graph rooted at 
  the center vertex, and \subref{fig:elem} an elementary part 
  $H^c \setminus \{v_2\}$ obtained from $H^{v_2}$, where the 
  gray vertex with its incident edge is deleted.}
  \label{fig:mccis-bbp}
\end{figure}
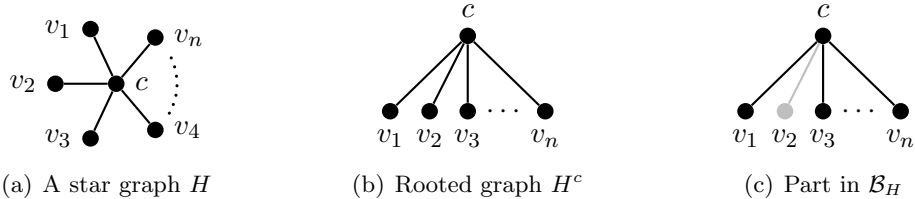

Let the two graphs $G$ and $H$ both be star graphs of order $n+1$, i.e., trees 
with all but one vertex of degree one as depicted in Figure~\ref{fig:star}. Since
trees are outerplanar, $G$ and $H$ are valid input graphs for BBP-MCS. 
The algorithm presented in~\cite{Schietgat2013} relies on a decomposition of the 
two input graphs into their parts.\footnote{%
The approach greatly simplifies for trees and we have shortened the required 
definitions accordingly. Please note that \cite[Algorithm 4 and Algorithm 3, lines 
11-18]{Schietgat2013} will not be required to solve the problem on trees.}
$\mathit{Parts}(T^r)$ of a rooted tree $T^r$ is recursively defined as follows 
\cite[Definitions~20,~23,~26]{Schietgat2013}.
\begin{enumerate}[(i)]
 \item $T^r\in \mathit{Parts}(T^r)$,
 \item if $P^p\in \mathit{Parts}(T^r)$ and $p$ is incident to exactly one edge $\{p,v\}$, then the graph $(P\setminus\{p\})^v$ is in $\mathit{Parts}(T^r)$,
 \item if $P^p\in \mathit{Parts}(T^r)$ and $p$ is incident to the edges $\{p,v_1\},\ldots,\{p,v_k\}$, $k \geq 2$, then for each edge $\{p,v_i\}$, $1\leq i \leq k$, the connected component of the graph $P^p\setminus\{\{p,v_j\} \mid j\not= i\}$ containing $p$ as root  is in $\mathit{Parts}(T^r)$.
 \end{enumerate}
For the first input graph $G$ an arbitrary root vertex $r$ is selected to define 
its parts.
Let $G$ be the star graph, $r$ its center vertex and let $L(G)$ denote its leaves,
then 
$$\mathit{Parts}(G^r) = 
\{ G^r \} \cup 
 \{(\{r,v\},\{\{r,v\}\})^r\mid v\in L(G)\} \cup 
\{ (\{v\}, \emptyset)^v \mid v \in L(G) \}.$$
The parts of the star graph are the graph itself, the subgraphs consisting 
of the individual edges and the subgraphs consisting of the leaves.
For the second input graph $H$, its parts are defined as 
$\mathit{Parts}^*(H)=\cup_{s\in V(H)}\mathit{Parts}(H^s)$~\cite[Definition~27]{Schietgat2013}.
Therefore,
\begin{align*}
\mathit{Parts}^*(H) =  
 &\{ H^s \mid s \in V(H)\} \ \cup \
 \{(e,\{e\})^c \mid e \in E(H)\} \ \cup\\
 &\{ (\{v\}, \emptyset)^v \mid v \in L(H) \} \ \cup \
 \underbrace{\{ H^c \setminus \{v\} \mid v \in L(H) \}}_{\mathcal{B}_H}, 
\end{align*}
where $c$ is the unique center vertex of $H$ and $\mathcal{B}_H$ the subgraphs
rooted at $c$ obtained by deleting a single leaf with its incident edge, cf. 
Figure~\ref{fig:elem}.

In order to solve the problem, a variant of BBP-MCS, which requires to map the 
root of one part to the root of the other, is solved for specific pairs of parts
denoted by $\mathit{Pairs}(G,H)$.
If the roots of both parts have multiple children, a matching problem between 
them must be solved. Such parts are referred to as \emph{compound-root graphs} 
and the parts associated with the children are \emph{elementary parts}, respectively~\cite{Schietgat2013}.
Note that this is the case for $G^r$ and all the parts in $\mathcal{B}_H$;
according to \cite[Definition~28]{Schietgat2013} we have $\{G^r\} \times \mathcal{B}_H \subseteq \mathit{Pairs}(G,H)$.
For each pair $(G^r, Q)$, $Q \in \mathcal{B}_H$, a weighted maximal matching 
instance is constructed, where the vertices correspond to the elementary parts 
of $G^r$ and $Q$~\cite[Algorithm 2, \textsc{RMCScompound}]{Schietgat2013}. 
The edge weights are determined by the solutions for pairs of smaller parts and 
depend on the possibly real-valued weights of vertex and edge labels of the 
common subgraph.
The number of elementary parts of $G^r$ is $n$,
the number of elementary parts of each $Q$ in $\mathcal{B}_H$ is $n-1$. Hence, 
each of these matching instances has $2n-1$ vertices and $n(n-1)$ edges and thus requires time $\mathcal{O}(n^3)$.
The number of such pairs is $|\{G^r\} \times \mathcal{B}_H| = n$. 
If each matching instance is solved separately, no better bound than 
$\mathcal{O}(n^4)$ on the total running time of the algorithm can be given and 
the analysis of \cite[Theorem 2]{Schietgat2013} is too optimistic.

Consequently, there must be an error in its proof:
The authors claim that every vertex $g \in V(G)$ and every vertex 
$h \in V(H)$ has at most $\deg(g)$ (resp. $\deg(h)$) elementary parts involved 
in a maximal matching. 
While this statement is correct the subsequent analysis does not take into 
account that there may be up to $\deg(h)$ matching instances of that size for a 
vertex $h \in V(H)$.
More precisely, the total time spent in \textsc{RMCScompound} for solving matching 
instances is claimed to be bounded by
\begin{equation}
 T_\text{comp} = \sum_{g \in V(G)} \sum_{h \in V(H)} T_\text{WMM}(\deg(g)+\deg(h)),
\end{equation}
where $T_\text{WMM}(k)$ is the running time for solving a weighted maximal matching
instance with $k$ vertices~\cite[p.~361]{Schietgat2013}.
Actually the procedure considers all pairs of compound-root graphs, where each 
pair leads to a matching instance containing one vertex for each of the associated 
elementary parts.
The counter example above shows that for a vertex $h \in V(H)$ there may be 
$\deg(h)$ compound-root graphs with root $h$, each with $\deg(h)-1$ 
elementary parts. In addition, there is one compound-root graph with root $h$
and $\deg(h)$ elementary parts.
Therefore, a correct upper bound is
\begin{equation}
 T_\text{comp}^\text{corrected} = \sum_{g \in V(G)} \sum_{h \in V(H)} (\deg(h)+1) \cdot T_\text{WMM}(\deg(g)+\deg(h)).
\end{equation}
In the counter example the degree of the center vertex is not bounded,
which leads to the additional factor of $n$ appearing in $T_\text{comp}^\text{corrected}$,
but not in $T_\text{comp}$.

\subsection{Exploiting the Structure of the Matching Instances}
The matching instances emerging for the counter 
example are closely related, since the symmetric difference of the elementary 
parts of $Q_1 \in \mathcal{B}_H$ and $Q_2 \in \mathcal{B}_H$ with $Q_1 \neq Q_2$ 
contains exactly two elements. It was recently shown that this fact can be exploited by solving 
groups of similar matching instances efficiently in one pass~\cite{Droschinsky2016}.
This technique was used to show that the maximum common subtree problem can be 
solved in time $\mathcal{O}(n^2\Delta)$ for trees of order $n$ with maximum 
degree $\Delta$, thus leading to $\mathcal{O}(n^3)$ worst case time. 
The same technique can be used to improve the running time of the BBP-MCS algorithm.

In~\cite{Droschinsky2016} the proposed maximum common subtree algorithm was
compared experimentally to the BBP-MCS algorithm of~\cite{Schietgat2013} using
the implementation provided by the authors. 
The running times reported for the BBP-MCS algorithm actually suggest a growth
of $\Omega(n^5)$ on star graphs.

\section{Violation of the Triangle Inequality}

Bunke and Shearer~\cite{Bunke1998} have shown that
\begin{equation}\label{eq:mcs-bs}
 d(G,H) = 1 - \frac{ |\textsc{Mcs}(G,H)| }{ \max\left\{\,|G|,\,|H|\,\right\} },
\end{equation}
where $|\textsc{Mcs}(G,H)|$ is the weight of a maximum common subgraph, is a metric 
and, in particular, fulfills the triangle inequality.
This was originally shown for a definition of the maximum common subgraph problem,
which requires common subgraphs to be induced and measures the weight of a graph
$G$ by $w(G)=|V(G)|$. Lins et al.~\cite{Lins2015} proved that Eq.~\eqref{eq:mcs-bs} 
also is a metric for the general (not necessarily induced) subgraph relation, 
where $w(G)=|V(G)|+|E(G)|$.
The article \cite{Schietgat2013} suggests that the weight of a BBP-MCS combined
with Eq.~\eqref{eq:mcs-bs} is a metric, too. We show that this is not the case.

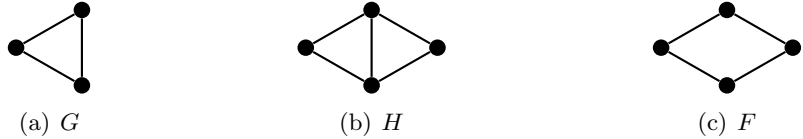
\begin{figure}
  \centering
  \null\hfill
  \subfigure[$G$]{\label{fig:mcces-bbp-non-metric:g}
\begin{tikzpicture}[auto]

\node[vertex] (u) at (0,0) {};
\node[vertex] (v) at (0,-1) {};
\node[vertex] (a) at (-0.866,-.5) {};

\draw[edge] (u) -- (v);
\draw[edge] (a) -- (v);
\draw[edge] (a) -- (u);

\end{tikzpicture}
  }\hfill
  \subfigure[$H$]{\label{fig:mcces-bbp-non-metric:h}
\begin{tikzpicture}[auto]

\node[vertex] (u) at (0,0) {};
\node[vertex] (v) at (0,-1) {};
\node[vertex] (a) at (0.866,-.5) {};
\node[vertex] (b) at (-0.866,-.5) {};

\draw[edge] (u) -- (v);
\draw[edge] (a) -- (v);
\draw[edge] (a) -- (u);
\draw[edge] (b) -- (v);
\draw[edge] (b) -- (u);

\end{tikzpicture}
  }\hfill
  \subfigure[$F$]{\label{fig:mcces-bbp-non-metric:f}
\begin{tikzpicture}[auto]

\node[vertex] (u) at (0,0) {};
\node[vertex] (v) at (0,-1) {};
\node[vertex] (a) at (0.866,-.5) {};
\node[vertex] (b) at (-0.866,-.5) {};

\draw[edge] (a) -- (v);
\draw[edge] (a) -- (u);
\draw[edge] (b) -- (v);
\draw[edge] (b) -- (u);

\end{tikzpicture}
  }
  \hfill\null
  \caption{Outerplanar graphs for which Eq.~\eqref{eq:mcs-bs} does not satisfy 
   the triangle inequality under BBP-MCS.}
  \label{fig:mcces-bbp-non-metric}
\end{figure}

Consider the example shown in Figure~\ref{fig:mcces-bbp-non-metric} and let the 
weight of a graph $G$ be defined as $w(G)=|V(G)|+|E(G)|$ following~\cite[Section 3.2, p.\,364]{Schietgat2013}.
Employing BBP-MCS, we obtain $|\textsc{Mcs}(G,H)|=6$, $|\textsc{Mcs}(H,F)|=8$ and 
$|\textsc{Mcs}(G,F)|=1$ and accordingly:
\begin{align*}
  \begin{array}{c|ccc}
    d & G & H & F \\
    \hline
    G & 0 & 1/3 & 7/8 \\
    H & 1/3 & 0 & 1/9 \\
    F & 7/8 & 1/9 & 0
  \end{array} 
\end{align*}
The triangle inequality is violated, since $d(G,F)>d(G,H)+d(H,F)$.
In general, the connectivity constraints imposed by BBP-MCS make it difficult
to derive a metric. For a more detailed discussion of this topic we refer the 
reader to~\cite[Section~3.6]{Kriege2015}.


\section*{Acknowledgements}
This work was supported by the German Research Foundation (DFG), priority 
programme ``Algorithms for Big Data'' (SPP 1736), project ``Graph-Based Methods for Rational Drug Design''.


\bibliographystyle{plainurl}
\bibliography{note_bbp_mcs.bib}

\end{document}